\documentclass{article}

\usepackage{amssymb,amsfonts,amsmath,stmaryrd}
\usepackage{cite,enumerate,float,indentfirst}
\usepackage{url}
\usepackage{color}
\usepackage{tikz}
\usetikzlibrary{arrows,snakes,backgrounds}

\def\be{\begin{eqnarray}}
\def\ee{\end{eqnarray}}
\def\nn{\nonumber}

\definecolor{red}{rgb}{1,0,0}
\definecolor{orange}{rgb}{1,0.5,0}
\definecolor{violet}{rgb}{0.7,0,1}

\hoffset= - 1.0in         % switch off for draft style
\begin{document}

%\hfill ??? \today

\hfill MIPT/TH-06/22

\hfill IITP/TH-07/22

\hfill ITEP/TH-09/22

\bigskip

\centerline{\Large{Machine learning of the well known things
}}

\bigskip

\centerline{ \bf V.Dolotin, A.Morozov, A.Popolitov}

\bigskip

\centerline{\it MIPT, ITEP \& IITP, Moscow, Russia}

\bigskip

\centerline{ABSTRACT}

\bigskip

{\footnotesize

Machine learning (ML) in its current form
implies that an answer to any problem
can be well approximated by a function of a very peculiar form:
a specially adjusted iteration of Heavyside theta-functions.
%$\theta(x) = \frac{1+{\rm sign}(x)}{2}$
%with specially adjusted coefficients.
It is natural to ask if
the answers to the questions, which we already know,
can be naturally represented in this form.
We provide elementary, still non-evident examples 
that this is indeed possible,
and suggest to look for a systematic reformulation 
of existing knowledge in a ML-consistent way.
Success or a failure of these attempts can shed light
on a variety of problems, both scientific and epistemological.  
%This note??? is a list of the simplest examples of this type.

%% ???
}

\bigskip

\bigskip

\section*{Prelude}

Machine learning (ML) \cite{MLbasics} nowadays penetrates
all the aspects of our everyday life
and promises to provide humanity with a new magic tool for every new problem
which we can happen to face.
Computer facilities do not seem to be restricted in any principal way and
can one day overcome the capacities of human mind --
which we can not improve in any way without interfering into the basic
biology of ourselves and giving up {\it human} for something else,
either biological or not.
This progress of artificial intelligence once again poses the question of
what {\it our} intelligence is and what are, if any, its potential advantages
over the artificial one.
This is one of eternal questions, however, today we can pose it more sharply.
This is because all the spectacular successes of modern ML
make use of the simplest possible steepest descent method --
which is tremendously far from much more advanced creations of human mind,
like physical theories and the dreams of the "theory of everything"
\cite{Gro,LL,GSW,UFN2},
which at the moment looks "relatively simple", still considerably more involved
than the offer of ML.
Can it be possible that just the most primitive optimization  method
allows us to learn everything, what human genius   achieved
during a number of centuries?

The answer seems obvious, at least to us.
And it is NO -- artificial intelligence, as we know it now,
is extremely efficient for solving optimisation problems,
i.e. in adjusting {\it parameters},
but is absolutely useless in finding the right algorithms,
i.e. understanding the {\it laws}, governing particular phenomena.
In other words, if we have no idea what is the {\it shape}
where {\it parameters} are to be adjusted,
efficiency of the adjustment is pretty low.
We can not generate new knowledge in this way,
only improve the quality of the existing one.
Laws of nature, as we know them, are {\it not} expressed through
Heavyside functions.

This seems to be a rather obvious idea,
still more and more papers appear where ML is "applied" to science,
we list just a few typical references from hep-th
%\cite{MLapps}
\cite{MLphysics}-\cite{BHHH} --
and an impression can mount that one can solve scientific problems in this way.
This short essay is to explain, why and in what sense this impression
is (can be?) wrong.

\section*{ML approach}

%\section{Machine learning}

Imagine that we want to associate a quantity $Z$ to a data $\vec X$
(recognize a cat in a photo, i.e. a special pattern in set of pixels),
i.e. find a function $g: \ \vec X \longrightarrow Z$.
For this purpose we use a given set of examples, labeled by $\alpha$,
when we postulate  the answer $Z^{\alpha}$ for the set $\vec X^{(\alpha)}$.
The idea of ML is to look for the function
$g: \ \vec X \longrightarrow Z$ by

(a) solving the minimal-square problem -- minimization of a functional like
\be
{\cal L}:= \sum_\alpha L^{(\alpha)} := \sum_\alpha \Big|Z^{(\alpha)} - G(\vec X^{(\alpha)}|W)\Big|^2
\label{Lfunc}
\ee

(b) for a special ansatz for $G(\vec X|W)$, when it is represented
as a combination of some $\sigma$-function
\be
G(\vec X|W) =   \sum_{i_1,i}  W_2^{i_1}\sigma(W_1^{i_1 i}X_i+b_1^{i_1 i} ) + +b_2^{i_1})
\label{G1}
\ee
Adjustment parameters are the coefficients $W$,
the minimum of $L$ is found by an iterative steepest descent method,
i.e. by solving the evolution equation
\be
\dot  W = - \frac{\partial {\cal L}}{\partial W}
\ee
for all the parameters $W$
and taking the limit of infinitely large time $t\longrightarrow \infty$.
What we get in the result is the function $\bar G(\vec X)\equiv G(\vec X|\bar W)$, 
which looks like (\ref{G1}),
i.e. is expressed through the $\sigma$-functions with particular (optimized) 
parameters $W = \bar W$.

%iteratively solve the minimal-square problem
%for the coefficients $W$:
%\be
%???L := \sum_\alpha || Z - \overbrace{\sum_{i,i_1,i_2,\ldots}\ldots
%\sigma\Big(W^{i_2i_1}\sigma(W^{i_1 i}X_i)\Big)}
%^{G\left(\vec X^{(\alpha)}\right)}
%\ee
Usually $\sigma(x)$ is a somehow smoothed version of the Heavyside $\theta(x)$,
like $\sigma(x) = {\rm arctanh}(kx)$,
but one can consider other choices as well.
%$Z^{(|alpha)}$ are the sample answers, associated with the data $\vec X^{|alpha}$
%and the task is to find an "analytical" solution to the map $G:\ \vec X \longrightarrow Z$.
An obvious generalization of the ansatz (\ref{G1}) is an iteration power of $\sigma$,
\be
G(\vec X|W) = \sum_{i,i_1,i_2,\ldots, i_m}  W_{m+1}^{i_m} \sigma\left(W_m^{i_{m}i_{m-1}}\sigma\Big( \ldots
\sigma\Big(W_2^{i_2i_1}\sigma(W_1^{i_1 i}X_i+b_1^{i_1 i})+b_2^{i_2 i_1})\Big)\ldots+b_{m+1}^{i_m}\Big)\right)
\label{Gm}
\ee
The number $m$ of $\sigma$-functions in this "approximation" is called
the {\it depth} (i.e. the number of layers) of the ML algorithm.
It can also involve a more sophisticated way to combine $\sigma$-functions,
which is called the {\it architecture} of the algorithm.
Also the cost-functional $L$ can be more complicated,
for example one does not need to make the arguments of $\sigma$ linear in $X$ --
already polynomials of higher degree can often lead to exponential increase in the efficiency 
of particular alrotihtms.
Finally, more elaborated setups, like unsupervised learning or learning of GANs (generator adversarial networks) are possible.
For simplicity, in this initiatory paper
we restrict consideration to the trivial multi-layer architecture
with the simplest Heavyside $\sigma(x) = \theta(x)$ and the simplest cost-functional (\ref{Lfunc}).
%The level is also assumed to be finite.???

In other words,
{\bf ML suggests to look for the answer in a somewhat special form (\ref{Gm})},
%(with some choice of parameters $W$, provided by optimization) --
which does not seem familiar from other branches of science.
Still, the apparent successes of machine learning imply that this shape
(iterative combination of $\sigma(x)$)
could be quite adequate.
Then we face a natural question:
{\bf if the answers, which we know, can be reformulated and brought to this form},
how easy it is and how natural is such reformulation?
These are the questions which we are going to address in this paper.
And its main content is just a set of elementary examples,
which can help to open the secret. 

In these examples we find an expression for the map $G(\vec X)$, known from ordinary science,
through the $\sigma$ (actually, Heavyside) functions -- and call it ${\cal G}(\vec X)$.
Then this implies an expression for the ML test function $G(\vec X|W)$ which can converge to
\be 
\bar G(\vec X) \equiv G(\vec X|W) = {\cal G}(\vec X) 
\ee
In other words, if the known answer can be "Heavysided",
then there is a ML process which has a chance to reproduce it.
If it can not, then there is a problem -- ML "answer" can not be true. 
An apparent resolution in this case is
instability of ML "answer" w.r.t. the variation of sample data.

\section*{Example: the zero of a function}

We have a function $y=F(x)$ and we wonder what is its zero.
The  answer to this question is
\be
{\cal G} \equiv
{\rm zero \ of \ }F =
\int_{dx} x F'(x) \delta\Big(F(x)\Big),
%=  \int_{dx} x   \theta'\Big(F(x)\Big)  ???
%%=  ????  \int_{dx}   \theta\left(  x\theta'\Big(F(x)\Big)\right) ???
\ee
provided that $F(x)$ only has \textit{one} zero, otherwise we get sum of
coordinates of all zeroes.

Conversion into the ML form in this case is immediate:
\be
%{\rm zero \ of \ }F
{\cal G}  
%= \int_{dx} x F'(x) \delta\Big(F(x)\Big)
=  \int_{dx} x \cdot  \theta'\Big(F(x)\Big)
%=  ????  \int_{dx}   \theta\left(  x\theta'\Big(F(x)\Big)\right) ???
\ee
The l.h.s. gives the right answer when the function has a single zero
(i.e. when the problem is well posed),
and the r.h.s. expresses this function in the ML form \eqref{G1}.

The discrete version of the formula, when integral is substituted by a sum
over the points of a, say, unit segment $x \longrightarrow k/N,\ \ k =1,\ldots, N$
says that
\be
%{\rm zero \ of \ }F
{\cal G}
= \sum_{k=1}^N  \frac{k}{N} \cdot \left\{\theta\left(F\left(\frac{x_k}{N}\right)\right) 
- \theta\left(F\left(\frac{x_{k-1}}{N}\right)\right)\right\}
= \sum_{k=1}^N  \frac{k}{N} \cdot \Big(\theta(f_{k}) - \theta(f_{k-1})\Big)
\label{ans1}
\ee
Heavyside function converts the values of the function $f=F(x)$ into just two:
\be
\theta(f) = \left\{\begin{array}{ccc}
1 & {\rm for} & f\geq 0 \\
0 & {\rm for} & f<0
\end{array}\right.
\ee
i.e. does not change unless the graph of a smooth function crosses zero.
The formula picks up the crossing and weights it appropriately.

Thus the ML cost-function $L$ in (\ref{Lfunc}) can be taken in the form
\be
L\{f\} = \left| \sum_{i=1}^N  \frac{i}{N} \cdot \Big(\theta(f_{i}) - \theta(f_{i-1})\Big)
- \sum_{i_1,i} W_2^{i_1} \sigma\Big(W_1^{i_1,i} f_i\Big)\right|^2
\ee
and one can expect that the steepest descent will converge to the point
\be \label{eq:zero-ans-1}
W_1^{i_1,i} = \delta_{i_1,i}    \nn \\
W_2^{i_1} = \frac{i_1}{N} - \frac{i_1+1}{N}
\ee
One can check that this is indeed the case, for $\sigma$ close enough to $\theta$.

\section*{An alternative architecture}

Though similar, yet slightly different in details, considerations
one can arrive at ``another'' shape (architecture) of the answer
to the zero-finding problem. Here they are.

%\bigskip
%\begin{verbatim}
%************************
%\end{verbatim}
%\bigskip

The first step in the (supervised) machine learning task is to provide computer
with a list of ``correct'' solutions to the problem (the so-called training set),
from which it would hopefully extract the general recipe. In this case training
data would consist of a large number of functions, together with their zeroes.
For simplicity let's assume that every function in the training set does, indeed, have exactly one zero.
Representation of the training data plays a big role: let us assume that every function is given
by its values on some mesh, say, $x_i = i$, $i = 1 \dots M$.

In such a case, the recipe for finding putative zeroes of a function would be:
\begin{itemize}
\item Loop through consequtive pairs of function values $(y_i, y_{i+1})$
\item If signs of the pair differ, we have a conjectural zero between $x = i$ and $x = i + 1$.
\end{itemize}

This recipe translates to the following machine-learning-type formula
\begin{align}
  x_{\text{zero}} = \text{AvgCoord}\left (
  \sum_{m=1}^M
  \sum_{l=1}^{2 M}
  \sum_{k=1}^{2 M}
  \sum_{i=1}^M
  \sigma\left(
  B^{(3)}_{m}
  +
  W^{(3)}_{m l}
  \sigma\left(
  B^{(2)}_{l}
  +
  W^{(2)}_{lk}
  \sigma\left(
  B^{(1)}_{k}
  +
  W^{(1)}_{ki}
  y_i
  \right)
  \right)
  \right)
  \right),
\end{align}
with
\begin{align} \label{eq:zero-ans-2}
  W^{(1)}_{ki} = & \delta_{k,i} - \delta_{k, i + M}, \ \ B^{(1)}_k = 0 \\ \notag
  W^{(2)}_{lk} = & \delta_{l,k} + \delta_{l, k + M+1} + \delta_{l + M - 1, k}, \ \ B^{(2)}_l = -1.5 \\ \notag
  W^{(3)}_{ml} = & \delta_{m,l} + \delta_{m + M, l}, \ \ B^{(3)}_m = 0 \\ \notag
  \\ \notag
  \text{AvgCoord(v)} = & \sum_{i=1}^M i \cdot v_i
\end{align}

Indeed, after application of the first network's layer $\sigma\left(B^{(1)}_{k} + W^{(1)}_{ki} y_i \right)$,
we have a $2M$-dimensional vector of (roughly) $1$'s and $0$'s, where on the first $1 \dots M$ positions
$1$ indicates that the corresponding $y_i$ is greater than zero and on the last $M+1 \dots 2 M$ positions
$1$ indicates that the corresponding $y_i$ is smaller than zero.

After application of the second layer $\sigma\left(B^{(2)}_{l} + W^{(2)}_{lk} \cdot \right)$
we have $2 M$-dimensional vector where on the $l$-th position is $1$ provided $y_l > 0$ and $y_{l+1} < 0$
and on $l+M$-th position is $1$ provided $y_{l-1} < 0$ and $y_{l} > 0$.
Such behaviour is ensured by the ``bias'' term $B^{(2)}$.

After application of the third layer we have $M$-dimensional vector where on $l$-th position
there is $1$ provided function changes sign in the vicinity of $l$.

Finally, the $\text{AvgCoord}$ calculates the average coordinate of a vector.

Already we see the problem/virtue of the ML approach.
Depending on our \textit{prior} knowledge about the problem, which
we introduce to our ML model in the form of architecture (ansatz)
the process converges to either \eqref{eq:zero-ans-1} or \eqref{eq:zero-ans-2},
\textit{in any case} confirming our presupposition.
We see here the potential sprouts of the \textit{unfalsifiability} problem:
the ML engine confirms (fulfills) both our initial plausible assumptions
(prophecies). Therefore \textit{additional} considerations are required to distinguish the right (most economical) hypothesis.

\section*{Multiplication problem}

Even more fundamental could be a ML approach in application to
addition and multiplication of integers.
%This example is a sort of  basic for distinguishing
%or, probably, defining(?) the human intellect.
Every operation allows to split a quantity into elementary objects.
For addition this building block is just $1$:
every natural number can be obtained by adding unities.
However the elementary objects for multiplication are numerous
and form a badly conceived set of the prime numbers --
the most mysterious of all phenomena of nature,
to which all the other mysteries are being slowly reduced
(through the study of adels, zeta-function, motives etc).
As one often says, "integers are the creatures of God" --
just as, probably, the human intelligence.
It is a crucial question if ML can learn the laws of multiplication
and the properties of primes.
%???Of course, one can reduce multiplication to addition by
%using the logarithm/exponential map --
%but what if we do not know it and need to invent?

In fact, addition and multiplication are considered as
the trivial (well understood)  inputs in (\ref{Gm}),
and one can think that there is no need to express them 
through Heavyside compositions. 
%-- what looks problematic for both $a+b$ and $ab$.
But there we assume addition and multiplication of real and complex
numbers, not integers.
For integers and rationals the problem of Heavysidization makes sense --
and can be easily resolved.
%and would be very interesting to address.

%\subsection*{Addition}

For addition of positive integers the possible formula is:
\be
a + b = \sum_{i=1}^\infty \theta(a-i) + \sum_{i=1}^\infty \theta(b-i)
\ee
where the two sums contain $a$ and $b$ unities respectively.
This is independent of the upper limit. which can be just some big number,
exceeding all relevant $a$ and $b$.
In ML language this solution implies that we use the test function
\be
G =\sum_iw_1^{i}\sigma(w_0^{i1}x_1+w_1^{i2}x_2+b_0^i)+b_1
\ee

%\subsection*{Multiplication}

%???We can approximate the multiplication map
%$\mathbb{R}\times\mathbb{R}\to\mathbb{R},\ (x_1,x_2)\mapsto y=x_1\cdot x_2$
%by the following level-2 network:
Similarly for multiplication
\be
a\cdot b = \sum_{i ,j}^\infty\theta\Big( \theta(a-i ) + \theta(b-j)-2 \Big)
\ee
Contributing are only the points $(i,j)$ where the argument of the external $\theta$
is non-negative, i.e. where both $\theta(a-i)$ and $\theta(b-j)$
are unities -- and there are exactly $a\cdot b$ such points.

Of course, the rules can be easily extended to rationals, at least with a somehow
limited denominator, but this is not a surprise --
the true border in number theory
is between rationals and other types of numbers, like algebraic or real.

For multiplication the problem looks more interesting.
A  possibility to simplify it
is to work in the binary system and
devise a Heaviside-based multiplication method for zeroes and unities.
%%?????For example,
%the simplest addition table
%\be
%0+0 = 00  \nn \\
%0+1 = 01  \nn \\
%1+0 = 01  \nn \\
%1+1 = 10  \nn
%\ee
%can be summarized as
%\be
%a+b =  \theta(a-1/2)\theta(b-1/2),  \theta ???
%\ee

\section*{Application to classification problems}

A standard ML task is classification: building a function on configuration space $X$ with values in a discrete set $Z$ of classes.

\subsection*{Example: 1D}
Take an example when $X=\mathbb{R}^1$ and $Z=\{0, 1\}$, i.e. we classify points of real line as belonging to 2 possible classes. The mapping which we want to approximate by our network looks as:
\be
g:X\to\left\{
\begin{array}{l}
0,\ x < 2 \\
1,\ 2\le x\le 3 \\
0,\ x>3
\end{array}\right.
\ee
i.e. points in the segment $[2,3]$ belong to the class $1$, and the rest to $0$.

Lets take a level-1 network. The output value has the general form
\be 
G=\sum_iw_1^i\sigma(w_0^ix + b_0^i)+b_1
\ee
If we take $\sigma$ to be a Heavyside function, 
then we can indeed express the well known answer in the following form:
\be 
{\cal G}(x)=\theta(x-2) - \theta(x-3) + 0 
\ee
This means that we can make the single layered network with 2 cells
which is going to converge to the values:
\be 
\bar w_0^1=\bar w_0^2=1,\ \bar b_0^1 =-2,\ \bar b_0^2 =-3
\\
\bar w_1^1=-\bar w_1^1=1
\ee

{\it One can see that this way we can  actually express 
any step function on the line, which solves the classification problem 
for the 1-dimensional configuration space $X$.}

\subsection*{Example: 2D}

%Take $X=\mathbb{R}^2$. 
For simplicity consider just a sector on the plane $X=\mathbb{R}^2$.
This means that the mapping we want to approximate by our network be
\be 
g:(x^1,x^2)\mapsto\left\{
\begin{array}{lcc}
1 &{\rm for} \  x^1, x^2>0\\
0 &   {\rm otherwise}
\end{array}\right.
\ee
being characteristic function of a sector.

Solution is provided by the two-layered network. 
In general for such network
\be
G = \sum_jw_2^j\sigma(\sum_iw_1^{ji}\sigma(w_0^{i1}x_1+w_0^{i2}x^2+b_0^i)+b_1^j)+b_2
\ee

The approximation  to our characteristic function gets exact for $\sigma=\theta$:
\be
{\cal G}(x_1,x_2) =-\theta\Big(\theta(-x_1)+\theta(-x_2)-1\Big)+1
%=g(x^1,x^2)
\ee
When both $x_1$ and $x_2$ are positive, the argument of external $\theta$ is $-1$, and $y=1$.
When any of $x_1$ or $x_2$ is negative the argument is $0$ or $1$, and $y=0$. 

Thus the relevant network is layer-2 with 2+1 cells and converges to 
\be
\bar w_0^{11}=\bar w_0^{22}=-1 \\
\bar w_1^{11}=\bar w_1^{12}=1 \\
\bar w_2^1=-1,\ \bar b_2=1
\ee

{\it Combination of sector characteristic functions may provide an approximation (with precision depending on the number of network cells) to any step function on the plane.}

Generalization  to higher dimensions seems straightforward:
to describe a sector in $D$ dimensions one needs a network of the level $D$.

\section*{Beyond classification}

Thus we learned how to find a root of $F(x)$
and demonstrated that the method is indeed the same as the ML
approach to classification problems.

But is it enough to obtain more explicit formulas, like
\be
\frac{-b \pm \sqrt{b^2 - 4ac}}{2}
\ee
for $F(x) = ax^2+bx+c$?

It is hard to give an immediate affirmative answer --
at least there is an apparent problem to address and examine.

\section*{The moral}

The above examples are instances of the archetypical scientific question:
solve the equation $F(x)=0$, and, as we just saw, it {\it can} be answered by ML.
Moreover, we identified the reason for this:
existence of the formula   (\ref{ans1})  for the answer ${\cal G}(x)$,
which is made from Heaviside functions $\theta\Big(F(x)\Big)$.
Thus we should ask ourselves -- if all available answers to all the
scientific question we know so far can be expressed as linear combinations
of iterated Heaviside functions?

If yes, then we can hope to adjust the coefficients of these linear combinations
by ML to the true values.

If not, ML will not provide us with a true answer.

If we use a convergent algorithm, some answer will be given --
the program will give a definite output.
It will perfectly match the sample data --
but it will have nothing to do with the {\it true} answer.
In particular it will not be stable under the change of a sample:
the adjusted coefficients will change dramatically when the sample
is modified, even slightly.

Given a problem, we can approach the emerging dilemma from two opposite sides:

\bigskip

{\bf
A) Find a representation to the answer through Heaviside functions
or prove that it does not exist for any finite number of iterations

\bigskip

B) Check the stability of the ML output under the change of the sample
}

\bigskip

Approach A) is a task for theorists,
and it is not quite simple as we hopefully demonstrated with above examples
for our standard ways of thinking.
B) can be handled by pure computer methods -- at least if we restrict
the number of layers in the network to be reasonably low.

The questions are:

(i) Can we express an answer to any scientific problem in the form of
Heaviside composition?

(ii) If yes, is there a universal way to convert any known answer to such form?

(iii) If not, what are the obstacles to such Heavysidization? 

This set of questions resemble those which led to constructive mathematics --
only now the restrictions seem to be much stronger.

Immediate bet for many would be that for most scientific problems
experiments with B)
will demonstrate that the ML method does not work.
The question is if we can provide such negative examples.
However, we explicitly demonstrated that the simplest questions --
addition and multiplication, finding zeroes of a function --
{\it can} be answered in this way, perhaps surprisingly.
The answer for the simplest case of a function zero was even not unique -- and this is a tip of another set of worries to be looked into.
The next task is to look at explicit formulas for algebraic numbers --
solutions to polynomial equations, both when explicit formulas are known,
and when only cohomological constructions {\it a la} non-linear algebra
\cite{GKZ,NLA} are available.

%And if we can explain, why the problem can not be reduced to the
%one in Example 1 --
%which seems to be the {\it mother of all scientific problems}
%that one could solve by any thinkable method.

We hope that this small essay can stimulate the search in both
directions A) and B) -- and help us to understand the limits of
machine learning and the fate of human mind in the times of
artificial intelligence.
{\bf Does science reduce to the steepest descent?}

%{\footnotesize
\section*{Reservation/Disclaimer}

To avoid possible misunderstanding, the problems raised in this note
concern the particular method of ML.
It should not be mixed with other kinds of Big Data analysis,
including the powerful induction method based on computer analysis
of various phenomena -- which we advocate and use for many years
\cite{NLA}-\cite{si}\footnote{
In fact some references in \cite{MLphysics}-\cite{BHHH}
also deal with {\it this} kind of approach,
rather than with the true ML.},
and which is getting increasingly important
with increase of computer capacities.
However, using computers to {\it strengthen}  human intellect
is not the same as to {\it substitute} it.
The question of the present note is whether the latter is also possible
and what should we do to understand if  ML provides us with this
possibility.
%}

\section*{Acknowledgements}

We are indebted  to A.Anokhin and S.Barseghian for useful discussions.

This work is partly supported by the Russian Science Foundation
(Grant No.21-12-00400).

\end{document}